# DEPOSITION OF THE STOICHIOMETRIC COATINGS BY REACTIVE MAGNETRON SPUTTERING

A. Sagalovych, S. Dudnik[1], V. Sagalovych

*[1]NSC "KIPT" (Kharkiv)*
*Ukraine*
*STC "Nanotechnology" (Kharkiv)*
*Ukraine*

The investigations of the reactive magnetron depositing of the stoichiometric coatings "metal-metalloid" were done. The dependences between sputtering parameters of a target and processes of plasmochemical formation on the surface of sample "metal-metalloid" and formations of coatings of the appropriate structure were investigated.
Experimental data on stoichiometric coatings AlN, $Al_2O_3$, TiN, $TiO_2$ is given. Features of reactive magnetron deposition and investigation results for obtaining of coatings with pregiven properties in particular for providing stability and controllability of coating deposition processes in time.

Исследованы стехиометрические покрытия "металл-металлоид", полученные реактивным магнетронным методом осаждения. Определены зависимости между параметрами распыления мишени и процессом плазмохимического осаждения на поверхность образца "металл-металлоид" при формировании покрытий соответствующей структуры. Приведены экспериментальные данные о стехиометрических покрытиях AlN, $Al_2O_3$, TiN, $TiO_2$. Определены особенности реактивного магнетронного напыления и получены результаты исследований относительно формирования покрытий с заранее заданными свойствами, в частности, для обеспечения устойчивости и управляемости процессов нанесения покрытий во времени.
**Ключевые слова**: метод реактивного магнетронного распыления, покрытия, оксид, нитрид, композит.

Досліджені стехіометричні покриття "метал-металоїд", які отримані реактивним магнетронним методом осадження. Визначено залежності між параметрами розпилення мішені й процесом плазмохімічного осадження на поверхню зразка "метал-металоїд" при формуванні покриттів відповідної структури. Наведено експериментальні дані про стехіометричні покриття AlN, $Al_2O_3$, TiN, $TiO_2$. Визначено особливості реактивного магнетронного напилювання та отримані результати досліджень щодо формування покриттів із заздалегідь заданими властивостями, зокрема, для забезпечення стійкості й керованості процесів нанесення покриттів у часі.
**Ключові слова**: метод реактивного магнетронного розпилення, покриття, оксид, нітрид, композит.

## INTRODUCTION

Plasma processes of obtaining and processing of various materials are the basis of many modern technologies. This is made possible by significant advances in both theoretical and experimental studies of these processes and the development of appropriate technological equipment [1].

Reactive methods for coatings deposition, allowing to obtain coating of the most extensive set of diverse properties, and in particular magnetron method, represent one of the groups of methods that are widely used in various technologies forming plasma coatings.

Depending on the functionality of coatings they must meet a number of requirements that can be satisfied by selecting the coating material, dopants, phase composition for multicomponent coatings, structure, density, adhesion, and other characteristics that define the service characteristics of coverage given conditions of its use. Therefore, in deciding obtaining coatings with predetermined properties are considered as aspects of selection of the coating, based on its known properties, its application possibilities in the form of coating in different ways, and depending on the coating conditions of its application. The latter is one of the main issues in the study of a method of coating conditions and its practical application. For vacuum plasma coating techniques in general and magnetron, including research processes of the coatings in terms of growth on the surface of the charged particles of various energies and densities, their influence on the properties of the obtained coatings are the most relevant. According to [2], the coatings obtained by magnetron sputtering depend on a sufficiently large number of parameters. They can be divided into parameters, associated with structural features of magnetron sputtering sources and spatial placement of substrate and its form, and parameters associated with temperature characteristics of the gas discharge (voltage, power, pressure, gas composition) and substrate (pretreatment, temperature, potential).

For coatings with predetermined properties we have to know their dependence on a combination of all the above parameters.

Today the question of application of magnetron deposition systems for metal plating in inert gas has been well studied. A lot of studies [3 – 13] to study the characteristics and level the impact on them design peculiarities magnetron deposition systems (strength and configuration of the magnetic field, geometry of the target), pressure, composition of the gas atmosphere and others.

And yet, despite the large number of studies on the effects of process parameters reactive magnetron sputtering on the properties of coatings, their comparison and the ability to playback characteristics to produce coatings with predetermined properties, in particular, to ensure proper handling and stability in time deposition processes that cause certain difficulties This is due to the fact that in most cases are common process parameters such as current, voltage, power level, the total pressure of the gas mixture and the ratio of the reaction gas to the inert gas in the mixture. They are not only closely related, but also to a large extent, depend on the characteristics of the magnetron, the total area of condensation, sputtering metal, pumping rate and some other factors. Therefore there is an urgent studies that have focused on studying the dependence of the properties of coatings obtained by reactive magnetron sputtering from a direct parameters that determine the conditions of their formation, in particular, the density and energy of charged particles in a general flow of material to the surface of the growth surfaces. Determining the influence of the ionic component in the flow of material to the substrate on the processes of formation of coatings by magnetron sputtering target, requires stabilization conditions of sputtering, which, in turn, specifies the density of ions and their proportion in the general flow sputtered matter. This motivates seek additional opportunities to change the density of ions directed to the substrate, without significant changes in the conditions of sputtering targets.

Taking into account diversity and complexity of the processes involved in the formation of coatings and ultimately influence their properties and their strong dependence on the design features of using equipment, processes of plasma formation of coatings studied not enough as the theoretical and experimental aspect to optimal selection conditions for their receipt of specified characteristics (composition, density, structural, orientation perfection, etc.).

The aim of this work is to study the conditions of obtaining coatings of different composition of oxide and nitride compounds stoichiometric reactive magnetron sputtering in terms of possibilities controlled changes in their composition and properties in different modes of magnetron deposition system and impact on the condition of such coatings processes of interaction target with the gas jet.

## ANALYSIS OF THE CONDITIONS OF REACTIVE MAGNETRON SPUTTERING IN OBTAINING COATINGS STOICHIOMETRIC COMPOSITION

Reactive coating methods allow to widely change the ratio of metal to metalloid coatings and thereby obtain them with wide variety of properties. A separate group of compounds covering a stoichiometric composition or close to it, because some properties have extreme values for films with precisely stoichiometric composition.

The presence of reactive gas in the working gas significantly affects the processes of excitation of plasma, density and ion energy, and the interaction with the target and the possibility of the formation of chemical compounds affect the condition and composition of the target surface, all mentioned above can lead to significant deviations from the normal run of current-voltage characteristics and other parameters peculiar to process of magnetron sputtering in an inert gas environment only [14 – 18]. In connection with this is complicated obtaining coatings with predetermined and stable characteristics and a number of studies of reactive magnetron sputtering linked with decision of question of stabilization modes deposition and reliable playback performance coatings in these processes. As noted in [2, 7, 14], in such cases, stabilization modes magnetron power supply current or power does not give satisfactory results in reproducing the characteristics of coatings, and the best results on deposition handling process is achieved by stabilizing the magnetron power supply voltage . By changing the ratio of flow of inert gas and reaction gas, the total pressure, pumping speed vacuum chamber can affect the shape of the current-voltage characteristics, location and severity of highs, location of areas of unstable modes magnetron deposition source and thus choose the most optimal in terms of controllability of the process of deposition, tuning ranges of magnetron deposition sources [2].

Reduction during operating time target thickness, configuration change of its surface by sputtering the target localization process, mainly in the area of the largest magnetic field, which is a common feature for magnetron sputtering systems both at work in inert gas and in mixtures with reactionary gases lead to changes in the rate of sputtering targets at other fixed parameters of the process and, consequently, to differences in the characteristics of the party coverages, which was obtained by seemingly constant process parameters [15]. In [19] this issue been studied in terms of opportunities to improve performance stability characteristics of coatings deposition reaction technologies and show that as operation of target, based on data on the drift voltage discharge at a fixed value of the discharge current in inert gas can enter adjustments to the mode deposition of coatings that provide sustainable reproduction of performance coatings AlN and $Al_2O_3$ by sputtering Al in the respective gas mixtures and reaction of inert gases. Descending areas of curves depending on the voltage on the partial pressure of the reaction gas corresponding discharge region of parameters where the coating is achieved by forming a desired composition.

One characteristic of the behavior of current-voltage relationships depending on the partial pressure of the reaction gas during magnetron sputtering targets in mixtures of inert gases and the reaction is a significant change in discharge voltage in a fairly narrow range of changes in the values of partial pressure of gas and histerezis on curves of voltage changes during reverse changing the partial pressure of the reaction gas or power in this region [15, 20, 21]. This is due to the processes of interaction between the reaction gas from the surface of the target, leading to the formation of compounds and, consequently, to a change in emission characteristics sputtering target that appropriately affect the characteristics of the discharge.

In terms of maximum performance reaction coating deposition, it is desirable to keep the process under the following values of the partial pressure of the reaction gas, which on the one hand, ensuring the formation of the desired coating, and, on the other hand, do not lead to significant "poisoning" of the target and a sharp decrease in the rate its sputtering. This corresponds with the initial sections of voltage drop discharge curves of its dependence on the partial pressure of the reaction gas. Since this region corresponds to a narrower range of parameters, then to control and maintain them at a given level are strict requirements that are difficult to satisfy because of the complex interrelationship between the partial pressure of the reaction gas, current and voltage level that can not be changed independently.

Another factor that complicates stabilization regime deposition coatings associated with local microarc discharges, the probability of which is increasing dramatically due to the formation on the surface of the target compounds with sharply different physical properties, such as dielectrics. Microarc discharges, besides leading to the emergence of droplet component in atomic flow to the substrate and, therefore, to uncontrolled changes in the composition and properties of the coating, causing severe disturbance "balance" in the interrelated process parameters, which may lead to a shift of the operating point sputtering magnetron system by histerezis.

One of the directions of improving handling reactive deposition associated with decreasing the influence of the reaction gas on the surface condition target by changing the ratio of the reaction gas to the inert gas that are served in the deposition chamber. Obviously, this can be achieved to some extent under resolving supply of inert gas into the zone of excitation level, i.e., to the target, and the reaction gas – to the substrate on which the coating is formed, which is done in many cases with distribution nozzles through which are presented separately reactive and inert gases and screens, limiting the space in the sputtering target zone and in the zone of condensation coating [20, 22]. This improves the conditions of the process and its stability.

Other direction in research is connected with attempts to find the most favorable conditions for sustainable management process through the optimal selection of the parameters that influence the course of discharge voltage curve depends on the partial pressure of the reaction gas, the value of the slope, length and position plot voltage drop on this curve.

In [15] analyzed the behavior of interconnected discharge characteristics by changing the discharge voltage, the surface condition of the target, as well as related changes in the partial pressure of the reaction gas in the chamber, the effective coefficient of sputtering target and the flux of metal atoms directed to substrate, with a constant flow of inert gases and reactive and speed of evacuation of the chamber. The dependence of current on voltage initially increases rapidly, then has a part of the recession, and then starts to increase again. Downturn on the voltage-current plot is an area that corresponds to the beginning and end of the exemption of the target surface film from metal compounds with reactive component gas mixture. This pattern of the relationship between voltage and current with the maximum, especially if it is narrow, cause instability mode discharge in the transition region, leading to an abrupt transition to one of the most extreme conditions the target surface – completely covered with active gas or completely free from film. This contributes to the corresponding change in the partial pressure increases when the target surface film begins to form compounds and, conversely, decreases when begins its destruction. If film formation on the target surface is mainly due to the active gas ions and not due to chemisorption of neutral gas particles, it is possible to smoothly control the degree of coverage of the target surface film through the optimal selection and voltage regulation on target. That is why in the process of reactive sputtering deposition to stabilize the regime recommended power magnetron to stabilize voltage, not current or power, as in the case of sputtering targets only in inert gas [15 – 17].

It is well known that in the case of unstable equilibrium of the system or process is easier to maintain their position or parameters around the equilibrium point in the oscillating rather than static. It is with this approach to solving the problem of stabilization of doing reactive sputtering processes another line of research was associated. Examples of this trend can be found in [21, 23]. Considering the progress curves current value and intensity of sputtering aluminum on the partial pressure of the reaction gas ($O_2$, $N_2$) in the area where there these curves have hysteresis, the authors identify links to the unstable equilibrium curves and, accordingly, the value of partial pressure, which define the boundaries of unstable conditions process. Outside unstable conditions are transitional areas that one side adjacent to the sputtering area clean target, and the other – to an area completely covered with film. If periodically changing current, then you can choose the conditions when the operating point on the current-voltage curve of the magnetron will alternately be located in an area that meets the clean target surface or completely covered, passing region of unstable equilibrium. Depending on the average current and its extreme values at the changes, frequency and modulation mode it will match some average value of the operating point on the current-voltage curve, and to obtain coverage of a given task is to ensure that this middle point quasistationary regime, on the one hand, was as close to a "theoretical" working point in a stationary mode of the magnetron, the other – the deviation of the concentration of elements in the process of film formation in this mode and the thickness of one period were not large enough to account for diffusion processes had the opportunity neutralize them on film thickness.

According to the existing experimental data and models of reactive magnetron sputtering [2 – 13] values for deposition, in which the ratio of metal to metalloid corresponding stoichiometric composition match with the boundary parameters early region of instability deposition assuming that the maximum ratio of metalloid to metal in the condensate meets their chemical formula, and the conditions of condensation coating on the substrate and other surfaces do not differ. In fact, there is always a region of homogeneity compound, resulting ratio metalloid to metal may exceed stoichiometric, and in addition, due to the deviation from equilibrium condensing conditions, this ratio may be even greater deviation from the stoichiometric ratio of elements in the compound. The result will be some shift parameters forming coatings of stoichiometric values that define the region of instability, and it is thus possible to obtain coating stoichiometric composition, without going into the region of parameter instability process.

The provisions of this unstable region of pressure and its value will be determined by a set of parameters, some of which are related to the hardware design process, i.e. magnetron characteristics and, in particular, the size of its target, the ratio of the total area of the vacuum chamber to the substrate and surface area, which will grow condensate pumping capacity of others – the options process control spraying, which primarily include the discharge current, working gas pressure and flow ratio of working and the reaction gases. Pressure of the working gas flows and value of working and reaction gases are mediated through parameters such as coefficients of sputtering targets, depending on the category of abuse, and which, in turn, depends on the above parameters in process control spraying. This includes temperature condensation.

## EXPERIMENTAL STUDIES OF THE COATINGS DEPOSITION PROCESSES

## EXPERIMENTAL EQUIPMENT AND METHODS OF RESEARCH

One of the factors that complicate the comparison and analysis of the study of the properties of coatings depending on process parameters reactive magnetron sputtering is closely interconnected conditions plasma chemical processes on the surface of the target and the substrate and the last is largely dependent on the equipment in which experiments are being conducted. Therefore, it is desirable to be able to influence their plasma substrate environment more independently than in conventional schemes magnetron sputtering. Therefore, conventional planar magnetron type deposition was supplemented by auxiliary excitation system RF discharge between the surface of the substrate and the target. RF discharge provides excitation discharge plasma with a relatively small discrepancy in the energies of ions (~ 3 ч 10 eV) with their average energy (10 ч 80 eV), which depends on the pressure and type of gas and does not depend on the power level. This makes it possible to control the plasma density by changing the discharge power and energy of ions, which are directed to substrates – by changing the accelerating potential applied to the latter. In the case of substrates from dielectric material or the deposition of dielectric films accelerating potential can be set by bias RF diode layer in the supply of high-frequency voltage to sample holder through matching device.

When choosing the optimal operating conditions for the process of magnetron sputtering is desirable to have the greatest possible range of possible values of the working gas pressure. A special interest is the area of low pressure (0,1 Pa), where film formation occurs under conditions of gas environment corresponding to molecular or near molecular nature of the movement of gas flows. And in this case RF discharge has advantages over many other ways excitation, with greater restrictions on working pressure gas environment, and in particular by its low values.

On fig. 1 shown a schematic diagram of the technological device that has been used in studies of the process of forming coatings using magnetron sputtering source.

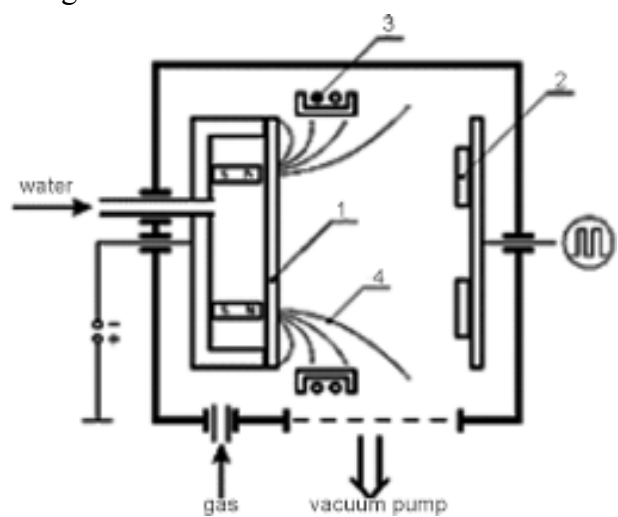

Fig. 1. Schematic diagram of the technological device: 1 – target; 2 – sample; 3 – inducer, 4 – magnetic field

Planar magnetron mounted on the side flanges of vacuum chamber installation. Magnetron target diameter 120 mm, maximum intensity radial component of the magnetic field at the target surface is 500 oersted.

For RF-discharge excitation between sample holder and target symmetrical axis of the target set double wound inductor in not weakly homogeneous magnetron magnetic field strength 15 – 20 oersted. Inductor is made of copper tubes with a diameter of 4 mm diameter coils 180 mm. Coils inductor protected by ceramic insulators to prevent the excitation of plasma intensive sputtering of the surface due to the large bias potential and contamination by products coatings sputtering.

The distance between the target and the sample holder is 150 mm, although in general may vary. Sample holder is electrically isolated from the vacuum chamber to be able to provide and regulate potential bias on substrate and control the total current flowing through it.

Power is provided from the magnetron power supply DC with strongly falling current-voltage characteristic power of 2 kW and open circuit voltage of 6 kV. RF power supplied to the coil through the matching device from serial generator UV-1 with an operating frequency of 13.56 MHz and a power of 1 kW.

To ensure controlled conditions of coating deposition in the initial period of their formation was additionally introduced valve, through which flow offset sputtered atoms from the target to the substrate during degassing and training target and entering at a given mode of the magnetron operation.

Substrates on which were deposited coating placed on sample holder at a distance of 90 mm from the surface sputtering target. Sample holder had drive rotation and was electrically isolated from the camera setup, which allowed it to apply controllable potential bias from DC source. As substrates were used foils from various metals thickness up to 0.5 mm in the form of rectangular strips of width up to 30 mm and length of 90 mm and plates of 25ч25 mm.

Before loading into the chamber substrates surface subjected to purification by degreasing petroleum ether, followed by washing in ultrasonic bath in distilled water. After drying the substrate fixed on sample holder and vacuum chamber was pumped to the pressure $(1 ч 2) \cdot 10^{-3}$ Pa, and then the camera filled with argon pressure $(1 ч 5)$ Pa and initiate glow discharge, which cathode was sample holder and anode - wall of cameras. Thus provided additional ion-plasma treatment sample surface prior to coating and the heating them to a temperature of 150 – 250 °C.

After a series of ion-plasma treatment reduced the bias voltage to a level of 0 ч 400 V, set given pressure in the chamber and composition of the working environment using SNA-2, set a given mode magnetron sputtering source and then open the shutter, which shut off the flow of substances deposition on the substrate.

In the process of coating controlled and maintained at a given level total pressure in the chamber, magnetron current, the bias voltage on the sample. Temperature of the substrates was monitored by a thermocouple mounted in sample holder.

Characteristics of the coatings determination were performed using X-ray diffraction techniques and phase analysis of X-ray diffractometer DRON-4, as well as scanning microanalyzer REMMA-200. Microhardness of coatings determined PMT-3 device at a load of 50 g.

## RESULTS OF EXPERIMENTS TO STUDY THE CONDITIONS OF COATINGS FORMATION

Current-voltage characteristics of the magnetron sputtering system at different pressures of the working gas determine the range of parameters of regulation and their sensitivity to changes in various areas of performance. This allows choose the field of sustainable modes of sputtering system, and the regulation that should be used to stabilize the mode depending on the position of the control point on a current-voltage curve.

On fig. 2 and 3 are given a family of current-voltage Characteristics of the sputtering magnetron system from copper and titanium target, respectively, at different pressures of the working gas.

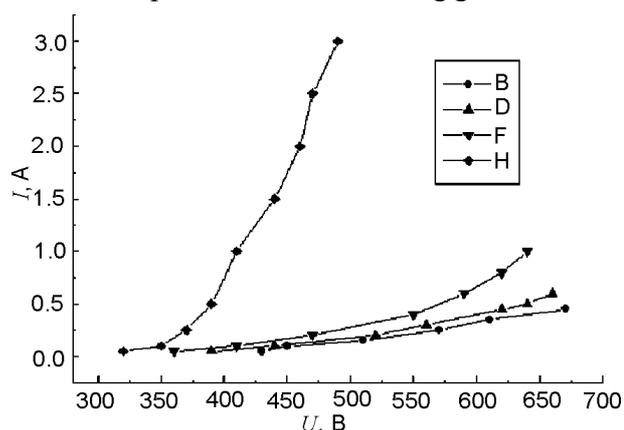

Fig. 2. Current-voltage characteristics of magnetron sputtering system when sputtering target of copper at different working gas pressure *P* (Pa): B – 0, 125; D – 0, 5; F – 1, 0; H – 2, 0.

Voltage of discharge extinction with minimal current value 0.05 A for copper was, depending on the pressure of argon, 320 – 430 V, for titanium – much less from 290 to 360 V. With increasing pressure the maximum value of the current increased and at a pressure of 1 Pa accounted for 1 A copper and titanium, about 2 A. Further increase in pressure in the vacuum chamber leads to a rapid drop in pumping speed of the diffusion pump (~ 1 Pa). To protect the diffusion pump overload and be able to raise the pressure in the vacuum chamber to values >1 Pa in vacuum system connecting the vacuum chamber with a diffusion pump valve is provided, by which you can reduce the passing aperture of vacuum system.

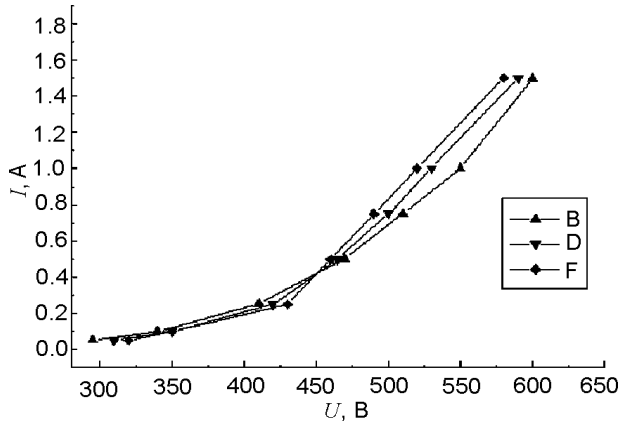

Fig. 3. Current-voltage characteristics of magnetron sputtering system when sputtering targets from titanium at different working gas pressure $P$ (Pa): B – 0, 25; D – 0, 5; F – 0, 8; H – 1, 0.

Using this valve was elevated pressure in the chamber to 2 Pa when sputtering copper target, thus achieved the maximum discharge current of 3 A. Thus, the experiments have established parameters constant region of sputtering system with appropriate power source when using argon from copper and titanium targets.

On fig. 4 and 5 are shown the current-voltage characteristics when sputtering titanium in argon when a pressure of 0.5 Pa and 0.8 Pa, respectively, and for different values of the partial pressure of nitrogen.

Introduction to argon nitrogen in an amount up to 4% did not significantly change the look of the current-voltage characteristics as compared with the characteristics of discharge extinction in pure argon. With regard to the current value, then the initial sections of the current-voltage characteristics, it slightly decreases and then increases with increasing voltage versus discharge in pure argon at the same voltage values. The point of intersection curves of current from voltage discharge in pure argon and mixtures of argon and nitrogen with an increase in the percentage of nitrogen is shifted toward larger values of voltage. Accordingly to this fact, the maximum difference in the magnitude of current in the initial and final regions of current-voltage characteristics increases with increasing nitrogen content in argon. Similar results on the effect of supplementation of nitrogen in argon on the dependence of cur-rent on voltage in comparison with the curve for pure argon when magnetron sputtering of silicon obtained in [14].

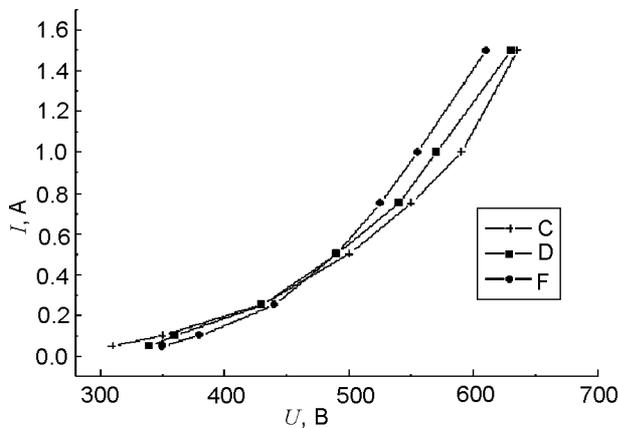

Fig. 4. Current-voltage characteristics of magnetron sputtering system when sputtering targets from titanium at different partial pressures of the reaction gas – nitrogen, $P$ (Pa): C – 0; D – 0, 01; F – 0, 02. Pressure of the working gas Ar – 0, 5 Pa.

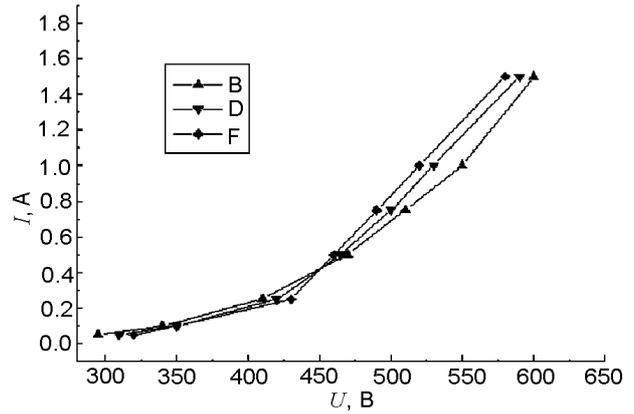

Fig. 5. Current-voltage characteristics of magnetron sputtering system when sputtering targets from titanium at different partial pressures of the reaction gas – nitrogen, $P$ (Pa): B – 0; D – 0, 01; F – 0, 02. Pressure of the working gas Ar – 0, 8 Pa.

Studies show that one of the parameters that are sensitive to changes of the target surface in reactive sputtering processes is decrease of discharge voltage when a fixed value of the discharge current.

On fig. 6 and 7 are given dependence discharge voltage value from the working gas mixture for different values of current and total pressure.

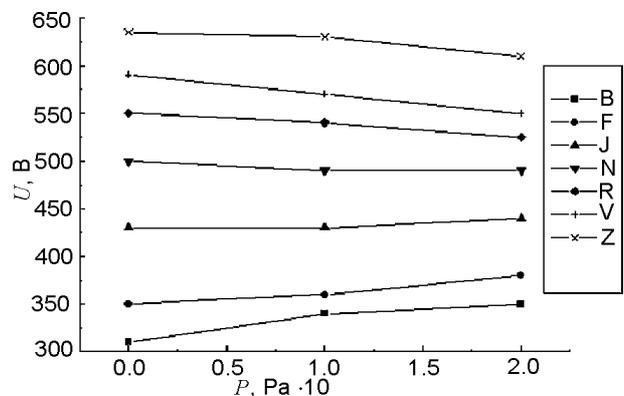

Fig. 6. Magnetron discharge voltage $U$ (V) vs the partial pressure of nitrogen $P$ (Pa) at different values stabilized discharge current $I$ (A): B – 0, 05; F – 0, 1; J – 0, 25; N – 0,5; R – 0, 75; V – 1, 0; Z – 1, 5. Pressure of the working gas of argon – 0, 5 Pa sputtering target-titanium.

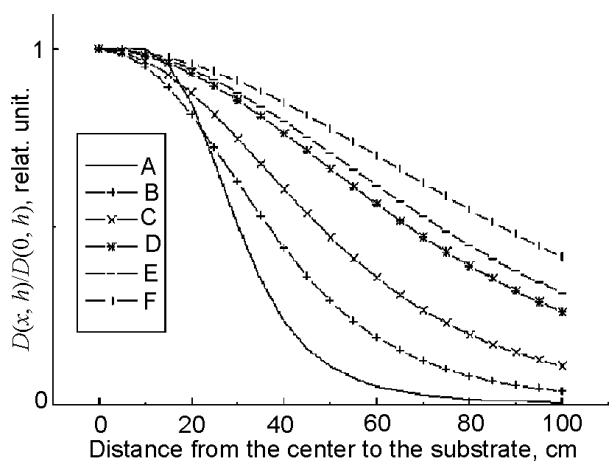

Fig. 7. Magnetron discharge voltage $U$ (V) vs the partial pressure of nitrogen $P$ (Pa) at different values stabilized discharge current $I$ (A): B – 0, 05; F – 0, 1; J – 0, 25; N – 0,5; R – 0, 75; V – 1, 0; Z – 1, 5. Pressure of the working gas of argon – 0, 8 Pa sputtering target-titanium.

It is seen that when a discharge current 0.5 A the slope of curves discharge voltage on the partial pressure of nitrogen is changing, the observed decrease in comparison with its value in pure argon. Upon reaching the discharge current 0.5 A and partial pressure of nitrogen 0,01 Pa appear microarcs, intensity and frequency of which increases with the discharge current and the partial pressure of nitrogen. Since the partial pressure of nitrogen ~ 1 Pa and discharge current ~ 0.5 A on the substrate formed coating of titanium compounds with nitrogen.

Thus, studies permit when developing technology of coating deposition magnetron sputtering identify several temperature ranges of their formation with qualitative differences, the temperature range below 100 °C, which formed a very intense film with a smooth surface and observed their cracking temperature range about 100 °C, where the film still retains mirror, but less intense and have only a few cracks. In the temperature range from 100 °C to 200 °C there is a transition from the formation of films with a smooth surface structured with a further reduction of tension and the complete absence of cracks. Increasing the temperature to 400 °C was accompanied by the formation of relief over the surface and signs of cut individual grains.

Experimental data on the conditions for obtaining coatings of different composition magnetron sputtering targets made of aluminum and titanium in the presence of nitrogen or oxygen are given in tabl. 1.

Summarizing the experimental results can be defined areas of parameters in which are formed mainly metallic or oxide coating, referring to the specific source magnetron sputtering. Thus, the metal

Table 1

Composition of coatings and parameters of a reactive magnetron sputtering

| № | Pressure (Pa) | | Discharge current, A (power, kW) | Coating growth speed, mkm/hour | Coating phase composition | Source |
|---|---|---|---|---|---|---|
| | Ar | Reaction gas | | | | |
| 1 | 0.5 | – | 0.5 | – | Ti | Own data |
| 2 | 0.5 | ≥ 0.015 | 0.5 | – | TiN | Own data |
| 3 | 0.15 | 0.03 | 0.5 | – | AlN | Own data |
| 4 | 0.23 | 0.013 – 0.217 | (4) | – | Al + AlN | [9] |
| 5 | 0.23 | 0.22 | (4) | 6.5 | AlN | Own data |
| 6 | 0.23 | 0.246 – 0.276 | (4) | – | AlN + Al | Own data |
| 7 | 0.532 | 0.6 – 0.65 | 10 | – | AlN | [6] |
| 8 | 0.532 | 0.56 – 0.6 | 10 | – | $Al_2O_3$ | Own data |
| 9 | 0.25 | 0.05 | 1.5 | – | AlN | [7] |
| 10 | 0.24 | <0.17 | (4, Ti) + (3, Al) | – | Al + TiN + AlN | [10] |
| 11 | 0.24 | 0.17 | (4, Ti) + (3, Al) | – | TiN + AlN | Own data |
| 12 | 0.24 | >0.17 | (4, Ti) + (3, Al) | – | TiN + AlN + Al + Ti | Own data |
| 13 | 0.2 | 0.25 | 2 | 0.54 | TiN | [4] |
| 14 | 0.4 | 0.1 | 2 | 2 | TiN | Own data |
| 15 | 0.14 | 0.03 | 0.8 | 0.29 | TiN | Own data |
| 16 | 1.15 | 0.18 | 1 | – | TiN | [11] |
| 17 | 0.3 | – | 0.5 | – | Ti | [12] |
| 18 | 0.3 | ≥ 0.05 | 0.5 | – | TiN | Own data |
| 19 | 0.3 | – | 0.5 | 5 | $MoS_2$ | Own data |
| 20 | 0.3 | ≥ 0.045 | 0.5 | – | $TiO_2$ | Own data |
| 21 | 0.53 | 0.027 | 0.65 | – | $TiO_2$ | [13] |

zirconium coating is virtually impossible to get at a current less than 0.07 A. The voltage must be > 500 V. The voltage is very sensitive to the state of the target surface in the process of reactive sputtering coating and its changes can serve as an indicator of the transition from formation of metal oxide coatings to. In accordance with this formation of oxide coatings was observed at voltages ranging from ~ 180 to 450 – 500 V and a current of 0.15 – 0.2 A, depending on the total pressure of a mixture of gases and their ratio, which changed from 0.1 to 0.3.

Despite the differences in current (power) magnetron discharge, the growth rate of coatings, differences in equipment used in the experiments, the pressure of the reaction gas in mixture with working gas (Ar), which was introduced into the deposition chamber, which corresponded to the formation of coatings with composition close to stoichiometric or close to it, is within one order of magnitude. Thus, for compounds AlN, $Al_2O_3$ these values are in the range $(0.3 – 6.5) \cdot 10^{-1}$ Pa, and connections TiN and $TiO_2$ – in the range $(0.15 – 2.5) \cdot 10^{-1}$ Pa.

Study of coatings composition depending on the pressure of the reaction gas during magnetron sputtering showed that the initially formed single-phase structure of metal coating that as growth pressure becomes a two-phase structure from composition of the metal – compound "metal-metalloid", and then, after reaching a certain pressure passes in a single phase from compound "metal-metalloid" (tabl. 1, pp. 1, 2, 5, 8). Further increase in pressure should lead to a transition in the region of instability coating and shift the operating point of the magnetron. Accordingly, the deposition rate decreases dramatically, but the composition of the coating does not have to undergo significant changes (tabl. 1, pp. 1, 8). This feature dependence on the pressure of the coating can be used to obtain guaranteed coatings with composition close to stoichiometric, although in this case, as already noted, lost its deposition rate. Therefore, the formation of coatings with composition close to stoichiometric, especially if you use significant power – to achieve high performance coating process, preference is often given to regimes that correspond to areas of parameters to limit their transition to unstable section.

Thus, we can define the basic tools that can be used to improve the handling of reactive magnetron sputtering. These are: a) separating supply reactive and inert gases, accordingly to the substrate and the target, and reduce the additionally of reactive gas on the surface of the target due to restrictive screens; b) increase the ratio of pumping speed to the speed of absorption of reactive gas by reducing the surface at which the binding of gas atomized metal or increase productivity pumping equipment installation; c) selection modes magnetron sputtering system with smoother dependencies between interconnected characteristics of discharge and partial pressure of the reaction gas in the chamber; d) stabilization sputtering process on the most appropriate parameter level (e.g., voltage, when the dominant mechanism of film formation on the target surface is interaction of ions with the active gas rather than adsorption of neutral gas); e) conduct the process of deposition in the quasistationary mode, i.e. its transfer mode oscillations around the unstable equilibrium point due to periodic changes in the rate of flow of the reaction gas or power settings magnetron sputtering system.

## CONCLUSION

Researches of parameters reactive magnetron coating deposition type "metal-metalloid" stoichiometric composition were carried out. These results explain the experimentally known fact proximity or phase parameters reactive magnetron sputtering coating stoichiometric with the region of instability sputtering target, and allow these areas to determine the parameters and their dependence on the characteristics of both hardware (performance of pumping, the total surface of the chamber and condensation surface coating, the surface of the target, etc.), and the parameters of deposition process (value flows of working and reactive gases, the discharge current).